\documentclass[twocolumn]{aastex63}

\newcommand{\mcrit}{$M_{\rm crit}$}

\usepackage{url}
\usepackage{breakcites}
\hypersetup{breaklinks=true}
\usepackage{savesym}
\savesymbol{tablenum}
\usepackage{siunitx}	
\restoresymbol{SIX}{tablenum}
\usepackage{soul}
\usepackage{pifont}
\usepackage{amsmath}
\usepackage{mathtools}
\submitjournal{ApJ}

\shorttitle{Critical halo mass for Pop III stars}
\shortauthors{Kulkarni et al.}
\graphicspath{{./}{figures/}}

\begin{document}

\title{The critical dark matter halo mass for Population III star formation: dependence on Lyman-Werner radiation, baryon-dark matter streaming velocity, and redshift}

\correspondingauthor{Mihir Kulkarni}
\email{mihir@astro.columbia.edu}

\author[0000-0002-9789-6653]{Mihir Kulkarni}
\affiliation{Department of Astronomy, Columbia University, 550 West 120th Street, New York, NY, 10027, U.S.A.}

\author[0000-0002-8365-0337]{Eli Visbal}
\affiliation{Department of Physics and Astronomy, University of Toledo, 2801 W. Bancroft Street, Toledo, OH, 43606, U.S.A}
\affiliation{Ritter Astrophysical Research Center, 2801 W. Bancroft Street, Toledo, OH, 43606, U.S.A}

\author[0000-0003-2630-9228]{Greg L. Bryan}
\affiliation{Department of Astronomy, Columbia University, 550 West 120th Street, New York, NY, 10027, U.S.A.}



\begin{abstract}
A critical dark matter halo mass (\mcrit) for Population III (Pop III) stars can be defined as the typical minimum halo mass that hosts sufficient cold dense gas required for the formation of the first stars. The presence of Lyman-Werner (UV) radiation, which can dissociate molecular hydrogen, and the baryon-dark matter streaming velocity both delay the formation of Pop III stars by increasing \mcrit. 
In this work, we constrain \mcrit\ as a function of Lyman-Werner flux (including self-shielding), baryon-dark matter streaming, and redshift using cosmological simulations with a large sample of halos utilizing the adaptive mesh refinement (AMR) code \textsc{enzo}. We provide a fit for \mcrit\ as a function of these quantities which we expect to be particularly useful for semi-analytical models of early galaxy formation. 
In addition, we find: (i) the measured redshift dependence of \mcrit\ in the absence of radiation or streaming is $(1+z)^{-1.58}$, consistent with a constant virial temperature; (ii) increasing the UV background increases \mcrit\ while steepening the redshift dependence, up to $(1+z)^{-5.7}$; (iii) baryon-dark matter streaming boosts \mcrit\ but flattens the dependence on redshift; (iv) the combination of the two effects is not simply multiplicative.
\end{abstract}



\section{Introduction} 
\label{sec:intro}

Population III (Pop III) stars, defined by their extremely low metallicities, are the first generation of stars to form after the big bang. In the context of the standard model of cosmology ($\Lambda$CDM), numerical simulations predict that the first Pop III stars formed in pristine dark matter ``minihalos'' with masses of $10^5-10^6 \si{M_{\odot}}$ \citep{Haiman96, Tegmark97, Machacek01, Abel2002, Bromm2002, Yoshida2003, Greif15}. Understanding when and where Pop III stars form is important given their role in early metal enrichment and the first stages of galaxy evolution. Observations of Pop III stars also have the potential to shed light on the particle nature of dark matter, since the abundance of minihalos is strongly reduced in certain models such as warm dark matter or fuzzy dark matter \citep{Oshea_norman_06_wdm, Sullivan18, Mocz19}. 

A number of upcoming observations have the potential to either directly detect Pop III stars or constrain their properties indirectly. Halos where Pop III star formation is suppressed at early times because of strong UV radiation can form Pop III clusters that can be detected if they are gravitationally lensed by foreground galaxy clusters \citep{Johnson10, Visbal16_cr7, Kulkarni19}.  Pop III stars in the mass range $140-250\, \si{M_{\odot}}$ may explode as pair-instability supernovae (PISN) which may be detected with the \emph{James Webb Space Telescope (JWST)} \citep{Whalen13, Hartwig18_sn}. Remnants of Pop III stars and extremely metal-poor stars can be studied using galactic archaeology to infer properties of Pop III stars and can be complemented by observations of high-redshift galaxies using upcoming thirty meter class telescopes such as the \emph{Extremely Large Telescope (ELT)} and the \emph{Thirty Meter Telescope (TMT)} \citep{Frebel15, Hartwig18_secondgen}. Additionally, the 21-cm signal from cosmological neutral hydrogen as well as line intensity mapping of He II \SI{1640}{\angstrom} can be used to constrain the properties of Pop III stars \citep{Fialkov13, Visbal15}. 

Detailed theoretical predictions are required to maximize the scientific return of these various observational probes of the first stars.
Semi-analytic models of Pop III star formation provide a computationally inexpensive method to predict when and where they form \citep[e.g.,][]{Magg18, Visbal18, Visbal20}. These semi-analytic models start with dark matter halo merger trees generated with cosmological N-body simulations or Monte Carlo methods based on the extended Press-Schechter formalism. Star formation is then followed in these halos with analytic prescriptions. 
One of the most crucial parameters in semi-analytic models of the first stars is the minimum dark matter halo mass required for Pop III stars formation, \mcrit. To make accurate observational predictions with semi-analytical models, it is important to understand how \mcrit\ evolves with various environmental effects. 

\mcrit\ corresponds to the typical halo mass where sufficient cold and dense gas is present to cause runaway collapse and star formation. The cooling of gas in pristine minihalos happens primarily through roto-vibrational transition lines of molecular hydrogen. As Pop III (and Pop II) stars form, they emit Lyman-Werner (LW) photons with energy in the $11.2 - 13.6$ eV range that can dissociate molecular hydrogen. As the cosmic star formation rate density (SFRD) increases, a corresponding background of LW radiation builds up. Halos in a region with a high LW background have their molecular hydrogen destroyed and are unable to cool efficiently. In such regions, the minimum critical halo mass required for cooling is increased \citep{Haiman96, Tegmark97, Machacek01, Oshea07, Wise-Abel07}. This effect of the LW background on \mcrit\ was first studied in hydrodynamical cosmological simulations with a statistical sample of halos in \cite{Machacek01}.

The critical mass for halos also depends on a phenomenon known as the baryon-dark matter streaming velocity, which was first pointed out by \cite{Tseliakhovich10}. Prior to cosmic recombination, baryons were coupled to radiation via Thompson scattering, whereas the dark matter and radiation were uncoupled. This creates a relative velocity between baryons and dark matter involving a quadratic term in the cosmological perturbation theory \citep{Tseliakhovich10}. The streaming velocity is coherent over several comoving Mpc and follows a Maxwell-Boltzmann distribution with an RMS velocity of $\sim 30$ km/s at recombination. It decreases with time as $\propto (1+z)$. In the regions with high streaming velocities, dark matter halos need to be more massive with deeper potentials for gas to cool and form stars, which results in an increase of the critical mass \citep{Greif12, Stacy12, Fialkov14_review, Schauer19}.

Apart from its dependence on LW radiation and baryon-dark matter streaming, many analytic models predict a redshift dependence for \mcrit, such that it increases with decreasing redshift \citep{Tegmark97, Haiman2000, Trenti09}, or assume a redshift dependence corresponding to a fixed virial temperature \citep{Visbal14}. Previous numerical works that estimate the critical mass using a statistical sample of halos have not seen a redshift evolution of \mcrit\  or see a constant \mcrit\ as a function of redshift. \citep{Machacek01, Schauer19}.

It has been nearly two decades since the publication of \cite{Machacek01}. 
Many things have improved since then such as increased computing power, modified reaction rates, as well as available prescriptions for self-shielding from LW radiation \citep{Wolcott11, Wolcott19}. In the past, the impacts of LW radiation and streaming velocity on $M_{\rm crit}$ have been studied separately \citep[although see][]{Schauer20} and the increase in $M_{\rm crit}$ when both effects are present has been assumed to be independent of each other and to be multiplicative \citep{Fialkov13, Visbal20}. With the improvement in computational capabilities, we aim to study the combined effects of LW radiation, dark matter baryons streaming and redshift on $M_{\rm crit}$ and provide an analytic fitting function that can be then used by semi-analytic models to make observational predictions.

This paper is structured as follows. In Section~\ref{sec:method}, we describe our simulation setup, details on the self-shielding prescriptions used, criteria used for calculating \mcrit, and the parameter space probed in this work. In Section~\ref{sec:results}, we present our results about \mcrit\  and its dependence on LW radiation, streaming velocity, and redshift. In Section~\ref{sec:discussion}, we provide a fit for $M_{\rm crit}(J_{\rm LW}, v_{\rm bc}, z)$ and discuss how our results compare with previous works. We summarize our main conclusions in Section~\ref{sec:summary}.

\section{Methodology}
\label{sec:method}

\subsection{Simulation setup}
We perform cosmological simulations using the adaptive mesh refinement (AMR) code \textsc{enzo} \citep{Enzo, Enzo_joss}. We use the energy conserving, spatially third-order accurate Piecewise Parabolic Method (PPM) for the hydro solver in all of our runs. \textsc{enzo} follows the non-equilibrium evolution of nine species (H, H$^+$, He, He$^+$, He$^{++}$, e$^-$, H$_2$, H$_2^+$ and H$^-$) and includes radiative processes. We included a uniform LW background radiation with an updated self-shielding prescription based on \cite{Wolcott19}. We updated the reaction rates in \textsc{enzo} with \cite{Glover15a, Glover15b}.

For all of our cosmological simulations, we assume a cosmology based on recent Planck observations \citep{Planck14}: $\Omega_m = 0.32$, $\Omega_\Lambda = 0.68$, $\Omega_b = 0.049$, $h = 0.67$ and $n_s = 0.96$. In order to simulate a larger statistical sample of minihalos, we increased the Planck normalization of the matter power spectrum to $\sigma_8 = 1$ (from $\sigma_8 = 0.83$). We discuss this choice in more detail below. We run most of our simulations with a box size of 0.5 h$^{-1}$ Mpc and initial resolution of $512^3$ cells and dark matter particles (corresponding to a particle mass of $\SI{100}{M_\odot}$). We refine cells into smaller cells based on their baryon mass, dark matter mass and Jeans length. We refine the cell if the baryon or dark matter density becomes higher than $4\times 2^{3 l}$ times the corresponding densities on the root grid ($512^3$) in the simulation where $l$ is the refinement level, meaning that cells with more mass than 4 times the initial dark matter particle mass or baryon equivalent will be refined. The Jeans length is resolved by at least 4 cells, and generally controls the refinement during the later parts of the baryonic collapse. We allow a maximum of six levels of refinement resulting in a minimum cell size of $\sim 1$ pc at z = 20. Once the maximum refinement level is reached, an artificial pressure is added to the smallest cells such that the Jeans length is always refined by 8 cells to avoid artificial fragmentation \citep{Truelove97,Machacek01}.
We use artificial pressure to follow a statistical sample of halos and to avoid slowing simulations with dense runaway gas collapse. 
For cases with high LW backgrounds or high streaming velocities, in order to have a sufficient number of massive halos, we run simulations with a larger box size of 1 h$^{-1}$ Mpc with a $512^3$ base grid. This corresponds to a dark matter particle mass of $\SI{800}{M_\odot}$. For these runs we use a maximum 7 AMR levels in order to maintain the same spatial resolution as our other simulations. See Section \ref{sec:discussion} for a further discussion on resolution tests.

We utilized the \textsc{rockstar} halo finder \citep{Behroozi13_rockstar} to identify halos and used $M_{200c}$ (the mass enclosed within a sphere of mean density 200 times the critical density) for our definition of the halo mass. We generated the merger trees using \textsc{consistent-trees} \citep{Behroozi13_trees} in order to trace the evolution of the halos. The analysis for this work was performed using \textsc{yt} \citep{yt} and \textsc{ytree} \citep{ytree}.

\subsection{Initial conditions}
We generate the initial conditions for our simulations using CICASS \citep{O'Leary12, McQuinn12}. A number of previous works have simulated the streaming velocity by simply adding a uniform velocity to the baryon velocity field at the starting redshift \citep{Stacy12, Greif12, Schauer19}. As pointed out by \cite{O'Leary12} and \cite{McQuinn12}, this ignores the evolution of the gas density as a result of the streaming velocity from cosmic recombination to the redshift of the initial conditions. CICASS calculates the effect of streaming self-consistently using perturbation theory and displaces baryons with respect to dark matter particles appropriately in the initial conditions -- an effect which was not considered in most previous works.

To increase the number of halos while maintaining a high spatial resolution, we run our simulations with an increased amplitude of density fluctuations (by setting $\sigma_8$ = 1). This change leads to a larger number of star-forming minihalos at earlier redshifts without increasing the simulation box size (which would reduce spatial resolution). This modification enables us to study a large statistical sample of star-forming minihalos without significantly changing the halo properties. We carried out a simulation with the standard $\sigma_8$ value and found that the cooling properties of individual halos closely matched those of the higher initial power spectrum.

\subsection{Criterion for $M_{\rm crit}$}
Our aim is to understand when and where Pop III stars form. Following the gas collapse to high densities in halos requires very high resolution and is typically done in zoom-in simulations where a region around the halo of interest has higher resolution to follow the density evolution \citep{Oshea07, Kulkarni19}. In a cosmological simulation suite like this, we cannot follow gas collapse in all halos because of limited computing resources. Hence, we use a criterion to identify halos that have cold and dense gas that would collapse soon and lead to star formation, but do not follow the process of runaway collapse to very high density.

We define a halo to have cold and dense gas if it has at least one cell on the highest refinement level with $T < 0.5 T_{\rm vir}$ and $n > \SI{100}{cm^{-3}}$. 
We find that most of the cells following this criterion have their cooling time shorter than the Hubble time and hence we expect them to undergo a runaway collapse even though it does not happen in our simulation because of artificial pressure.     
We then examine the halos in each simulation output (corresponding to a particular choice of LW background, streaming and redshift) and try to determine \mcrit\ for that output. As shown in \cite{Machacek01}, there is a general trend such that increasing mass halos are more likely to have cold-dense gas; however, this trend is not perfect and so there is no unique way to determine \mcrit.  We tried various methods to define this quantity including the fitting function used in \cite{Machacek01} which also takes into account the total mass of cold dense gas in the halo. However, this method becomes unreliable when the number of halos is small and does not provide a reliable estimate of the uncertainty on the value of \mcrit. 

\begin{figure}
    \centering
    \includegraphics[width=0.45\textwidth]{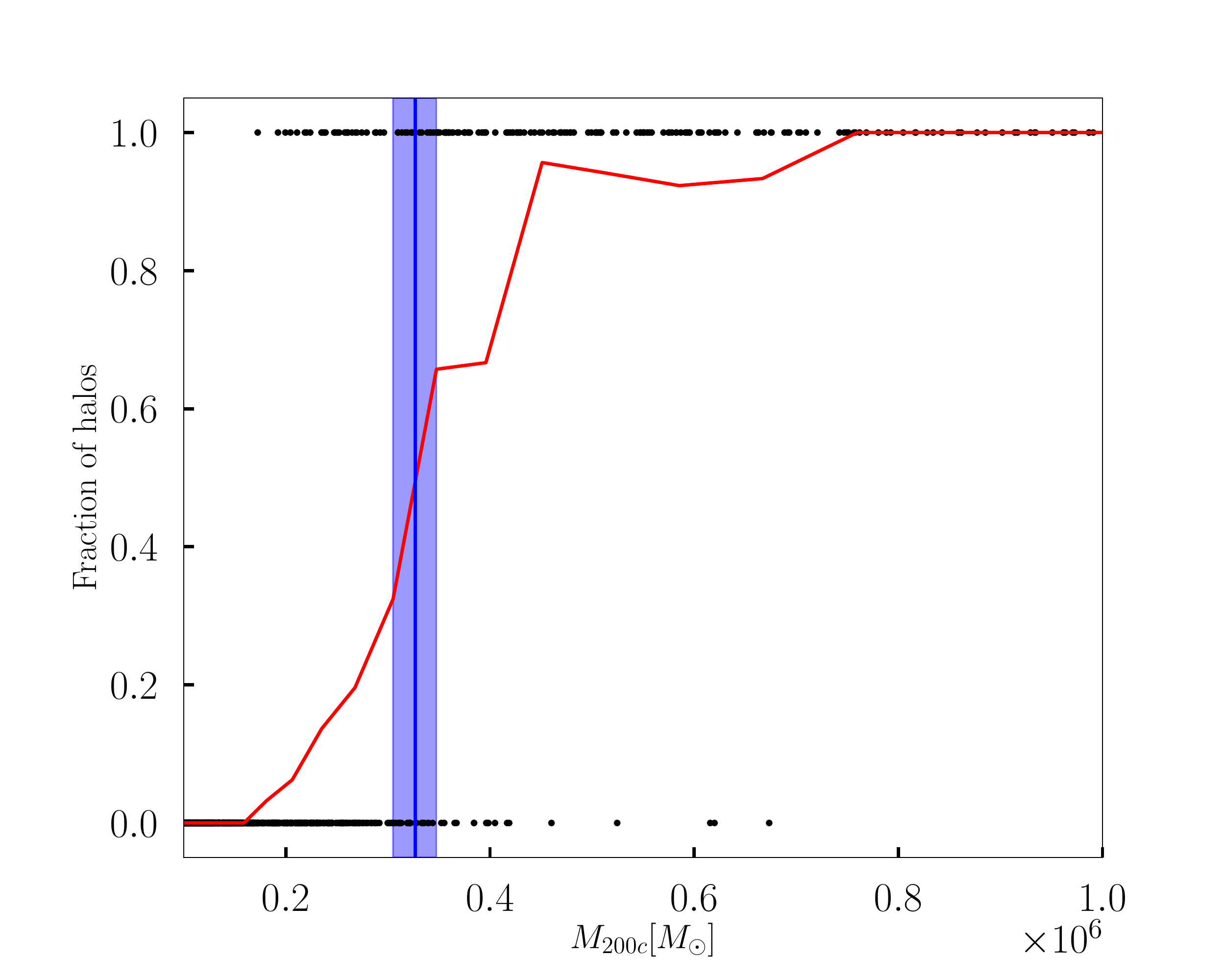}
    \caption{The scatter in halos that have cold dense gas. This demonstrates the method we used to identify \mcrit\ for halos from a run with $J_{\rm LW} = 0$ and $v_{\rm bc} = 0$ at $z = 15$. The black dots denote halo masses and their position on the y-axis indicates whether the halos have (1), or do not have (0) cold, dense ($T < 0.5 T_{\rm vir}$, $n > \SI{100}{cm^{-3}}$) gas. The red line shows the fraction of halos with cold-dense gas in each mass bin. The bin size corresponds to the smallest bin that can have a monotonically increasing red curve between $y = 0.25$ and $y = 0.75$. The critical mass corresponds to the mass bin where half of the halos in that bin have cold dense gas (shown in blue) and we use the bin size as an estimate of the uncertainty of this measurement.}
    \label{fig:criterion}
\end{figure}

Instead we adopt the following approach, as illustrated in Figure~\ref{fig:criterion}, to account for the scatter. Figure~\ref{fig:criterion} shows the halos with and without cold dense gas for $J_{\rm LW} = 0$ and $v_{\rm bc} = 0$ at $z = 15$. We bin halos into log-spaced mass bins and use the smallest bin size such that the fraction of halos that have cold-dense gas is monotonically increasing in the range of fractions from 0.25 to 0.75. This technique automatically adjusts for the number of halos since the bin sizes are naturally smaller for more halos (resulting in a more precise measurement of \mcrit) and larger for outputs with fewer halos (resulting in a less precise but more robust determination of \mcrit). As can be seen in Figure~\ref{fig:criterion}, we find a large scatter in halo masses for which a halo has cold-dense gas. We define the critical mass corresponding to the mass bin where half of the halos in that mass bin have cold-dense gas and we use the bin size as the uncertainty (error bar) on \mcrit. When we have fewer halos with cold dense gas at a given redshift, we need to use larger mass bins to satisfy the criterion, resulting in a larger uncertainty. Using slightly different cutoffs for the temperature and density does not change \mcrit\ significantly.
We also quantify the scatter around \mcrit\ in Section \ref{subsec:scatter}. 

We seek to characterize the simultaneous dependence of \mcrit\ on redshift, LW flux, and magnitude of the baryon-dark matter streaming velocity. 
We accomplish this by taking simulation snapshots at $z =$ 30, 27, 25, 24, 23, 22, 21, 20, 19, 18, 17, 16, and 15 for a number of different runs with various combinations of LW flux and streaming velocity. These include 4 LW backgrounds: 0, 1, 10 and 30, in units of $J_{21}$ where $J_{21} = \SI{e-21}{erg ~ s^{-1} cm^{-2} Hz^{-1} Sr^{-1}}$.
LW backgrounds of up to $J_{21}$ are expected to be common, however cases with 10 and 30 $J_{21}$ would correspond to regions with high LW radiation, possibly from a nearby source; these high values are still smaller than $J_{\rm crit}$ \citep{Shang10,Wolcott17}, the value required to dissociate molecular hydrogen to a sufficient extent that the gas stays warm ($T \sim 10^4$ K) throughout the collapse \citep[e.g.,][]{Ahn09}. 

For baryon-dark matter streaming, we include three cases corresponding to 0 km/s, 30 km/s ($1\sigma$) and 60 km/s ($2\sigma$), at recombination. Table~\ref{table:parameters} shows a grid of parameters of LW backgrounds and dark matter-baryon streaming velocities used in our simulation suite. This grid is chosen to study the effect of the LW background and streaming independently, as well as to test the assumption of independence when both processes are present, something that has been assumed in previous works \citep[e.g.,][]{Fialkov13}.

\begin{table}
\begin{center}
\vspace{0.1in}
\begin{tabular}{ |c|c|c|c|c| } 
 \hline
   & $J_{21} = 0$ & $J_{21} = 1$ & $J_{21} = 10$ & $J_{21} = 30$ \\ 
  \hline
 $v_{\rm bc} = 0$ & \checkmark & \checkmark & \checkmark & \checkmark \\ 
 \hline
 $v_{\rm bc} = 1 \sigma$ & \checkmark & \checkmark & \checkmark &  \\ 
 \hline
 $v_{\rm bc} = 2 \sigma$ & \checkmark & \checkmark & & \\
 \hline
\end{tabular}
\caption{The LW background and dark matter-baryon streaming parameters used in our simulation suite. A check indicates that we ran that combination of parameters. }
\label{table:parameters}
\end{center}
\end{table}

\section{Results}
\label{sec:results}

In this section we describe our results for \mcrit\ and its dependence on redshift, LW background intensity and the magnitude of the baryon-dark matter streaming velocity. In the following subsections, we describe the dependence of \mcrit\ on redshift with LW radiation in the absence of streaming, and streaming in the absence of LW radiation. Finally, we present the effect on \mcrit\ when a LW background and streaming are simultaneously included. 

\subsection{LW flux, no streaming velocities}
\label{subsec:LW}

\begin{figure}
    \centering
    \includegraphics[width=0.5\textwidth]{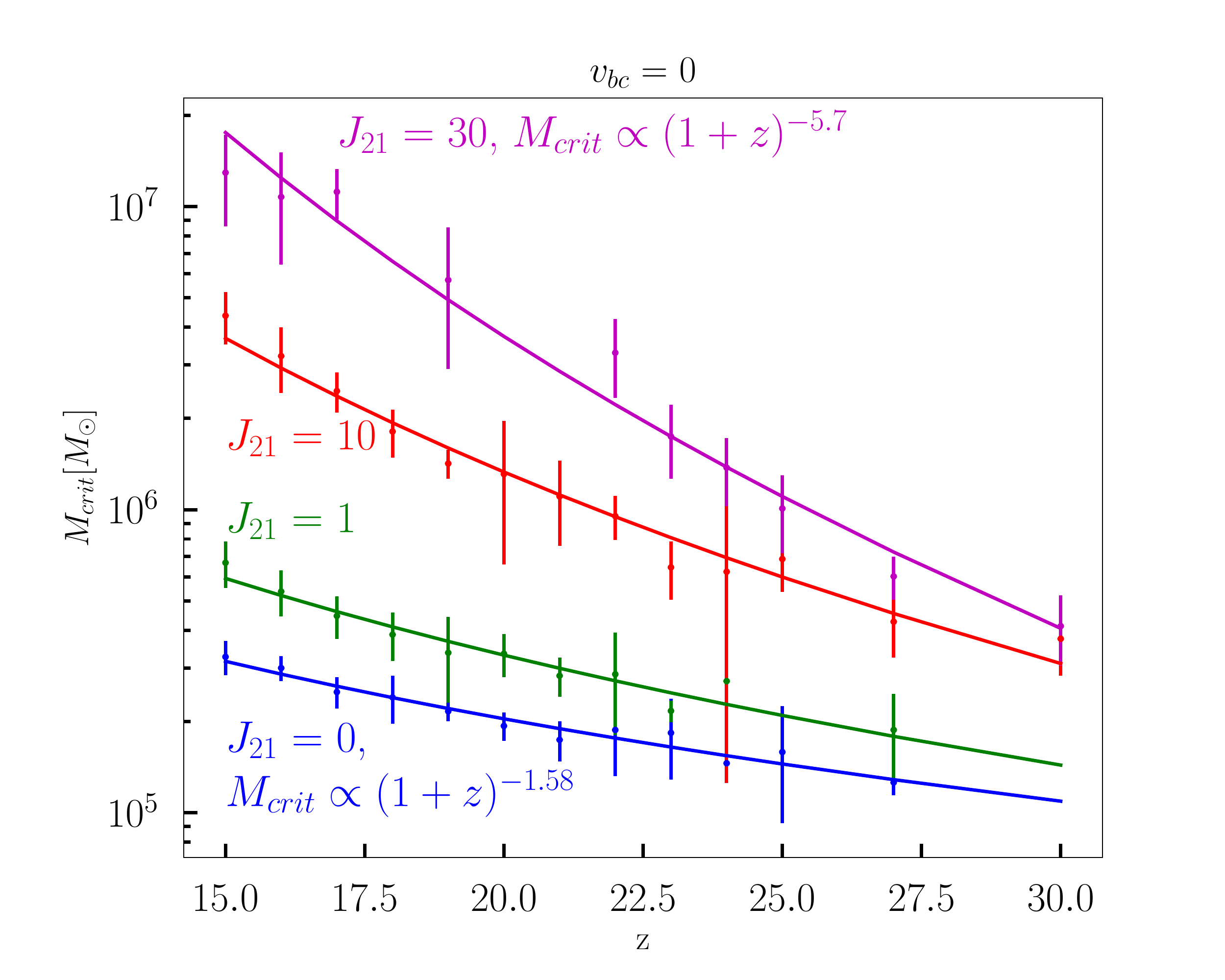}
    \caption{The value of $M_{\rm crit}$ as measured from a set of simulations with varying LW background as a function of redshift for a region with no dark matter-baryon streaming. The four lines correspond to LW backgrounds of 0, 1, 10 and 30 $J_{21}$ respectively. $M_{\rm crit}$ increases with decreasing redshift. The exponent of (1+$z$) changes from -1.6 to -2.1, -3.7 and -5.7 for LW backgrounds going from 0 to 1, 10 and 30 $J_{21}$.}
    \label{fig:v0_LWs}
\end{figure}

Figure~\ref{fig:v0_LWs} shows \mcrit\ as a function of redshift for four different LW backgrounds of $J_{\rm LW} =$ 0, 1, 10 and 30 $J_{21}$. For each redshift, \mcrit\ is measured, along with an estimate of the uncertainty, as described in Section~\ref{sec:method}. We will begin by discussing the case of $J_{21} = 0$ (blue curve at the bottom). We observe that \mcrit\ increases with decreasing redshift. This can be fit as a power law with $M_{\rm crit} \propto (1+z)^{-1.58}$ as can be seen from the top row of Table \ref{tab:fits}. The redshift dependence of \mcrit\ has been previously proposed in analytical models \citep{Haiman96, Tegmark97, Trenti09}, and has often been assumed to correspond to a fixed virial temperature and therefore vary as $M_{\rm crit} \propto (1+z)^{-1.5}$ \citep{Visbal14}; however this dependence has not been detected in previous simulations with a statistical sample of halos \citep{Machacek01, Schauer19}. Here we find a dependence consistent with this redshift evolution.
This dependence is relatively simple to understand: if the gas temperature in the halo is approximately equal to the virial temperature before runaway cooling, then, since both the $H_2$ formation rate and the cooling rate depend most sensitively on temperature, efficient cooling should depend mostly on the virial temperature. Our results nicely confirm that picture.

Figure~\ref{fig:v0_LWs} shows that, as the LW background increases from $J_{\rm LW} = 0$ to $J_{\rm LW}$ = 1, 10 and 30 $J_{21}$, then \mcrit\ also increases. 
Boosting the LW radiation dissociates molecular hydrogen, hence halos need to be more massive to host cold-dense gas in the presence of a high LW radiative flux. 
This increase in \mcrit\ with LW background can also be seen from Table \ref{tab:fits}. The third column ($M_{z=20}$) denotes \mcrit\ at $z = 20$ for a fit assuming a power law for redshift dependence. $M_{z=20}$ increases from $\num{2.04e5} \pm \num{3.79e3} \si{M_{\odot}}$ for $J_{\rm LW} = 0$ to $\num{3.73e6} \pm \num{2.73e5} \si{M_{\odot}}$ for $J_{\rm LW} = 30 J_{21}$.
At $z = 15$, we see an increase in $M_{\rm crit}$ by a factor of $2-3$ when going from $J_{\rm LW}=0$ to $J_{\rm LW} = J_{21}$, whereas that increase was by a factor of $15-20$ in some previous works \citep[e.g.,][]{Machacek01, Oshea07, Visbal14}.
The primary reason for this difference with those papers is the inclusion of the improved self-shielding prescription from \cite{Wolcott19}; note that no self-shielding was included in \cite{Machacek01} or \cite{Oshea07}. 
The effect of self-shielding will be discussed in more detail in Section~\ref{sec:discussion}. 
For the highest LW background considered here (30 $J_{21}$), we see an increase in $M_{\rm crit}$ by nearly two orders of magnitude, increasing from \SI{3e5}{M_{\odot}} to \SI{2e7}{M_{\odot}} at $z = 15$, as shown in Figure~\ref{fig:v0_LWs}.

Apart from seeing a redshift dependence on \mcrit\ for $J_{\rm LW} = 0$, we also find that the redshift dependence of \mcrit\ gets steeper with increasing LW flux, as can be seen in Figure~\ref{fig:v0_LWs} and Table~\ref{tab:fits}.
The exponent of the $z$-dependence changes from $-1.58$ when $J_{\rm LW}=0$ to $-2.14$, $-3.74$ and $-5.70$ for $J_{\rm LW}$ of 1, 10 and 30 $J_{21}$, respectively as can be seen from the first 4 rows of Table~\ref{tab:fits}. This means that the increase in \mcrit\ due to the LW background is more prominent at lower redshifts than at higher redshifts. \mcrit\ for all the redshifts can be found in the accompanying file.

We can better understand this redshift dependence by examining the molecular hydrogen content in halos, which is discussed in more detail in Section~\ref{subsec:gas_properties}.
We describe a simple analytic model to understand the steeper $z$-dependence in presence of LW background in Section~\ref{subsec:model}.

\begin{table*}[t]
    \centering
    \begin{tabular}{|c|c|c|c|c|c|}
         \hline
         $J_{\rm LW} (J_{21})$ & $v_{\rm bc}$ (km/s) & $M_{z=20} (M_{\odot})$ & $\alpha$ & $Q_1 (M_{z=20} (M_{\odot}))$ & $Q_3 (M_{z=20} (M_{\odot}))$ \\
         \hline
         0 & 0 &\num{2.05e5} $\pm$ \num{4.17e3} & 1.58 $\pm$ 0.13 & \num{1.76e5} & \num{2.81e5} \\
         1 & 0 &\num{3.31e5} $\pm$ \num{9.78e3} & 2.14 $\pm$ 0.18 & \num{2.76e5} & \num{4.33e5}\\
         10 & 0 &\num{1.33e6} $\pm$ \num{5.03e4} & 3.74 $\pm$ 0.19 & \num{9.80e5} & \num{1.90e6}\\
         30 & 0 &\num{3.73e6} $\pm$ \num{2.73e5} &  5.70 $\pm$ 0.32 & \num{2.98e6} & \num{4.26e6}\\
         0 & 30 ($1\sigma$) &\num{6.71e5} $\pm$ \num{1.22e4} & 1.05 $\pm$ 0.09  & \num{5.28e5} & \num{8.80e5}\\
         1 & 30 ($1\sigma$)&\num{1.01e6} $\pm$ \num{2.11e4} & 2.00 $\pm$ 0.11 & \num{8.29e5} & \num{1.16e6}\\
         10 & 30 ($1\sigma$)&\num{2.90e6} $\pm$ \num{2.06e5} & 4.31 $\pm$ 0.41 & \num{1.94e6} & \num{3.74e6}\\
         0 & 60 ($2\sigma$)&\num{1.84e6} $\pm$ \num{2.05e5} & 1.06 $\pm$ 0.59 & \num{1.26e6} & \num{2.37e6}\\
         1 & 60 ($2\sigma$)&\num{2.81e6} $\pm$ \num{1.17e5} & 1.16 $\pm$ 0.24 & \num{2.39e6} & \num{3.26e6}\\
         \hline
    \end{tabular}
    \caption{The fit parameters for redshift evolution fit as $M_{\rm crit} = M_{20} (1+z/20)^{-\alpha}$ for all the combinations of LW background and streaming velocities used in the simulations. The fifth and sixth columns show the first and third quartiles respectively corresponding to masses where 25\% and 75\% of the halos at that mass have cold-dense gas. These quantities can be used to estimate the scatter around \mcrit.}
    \label{tab:fits}
\end{table*}

\subsection{Streaming velocities, no LW flux}

\begin{figure}
    \centering
    \includegraphics[width=0.5\textwidth]{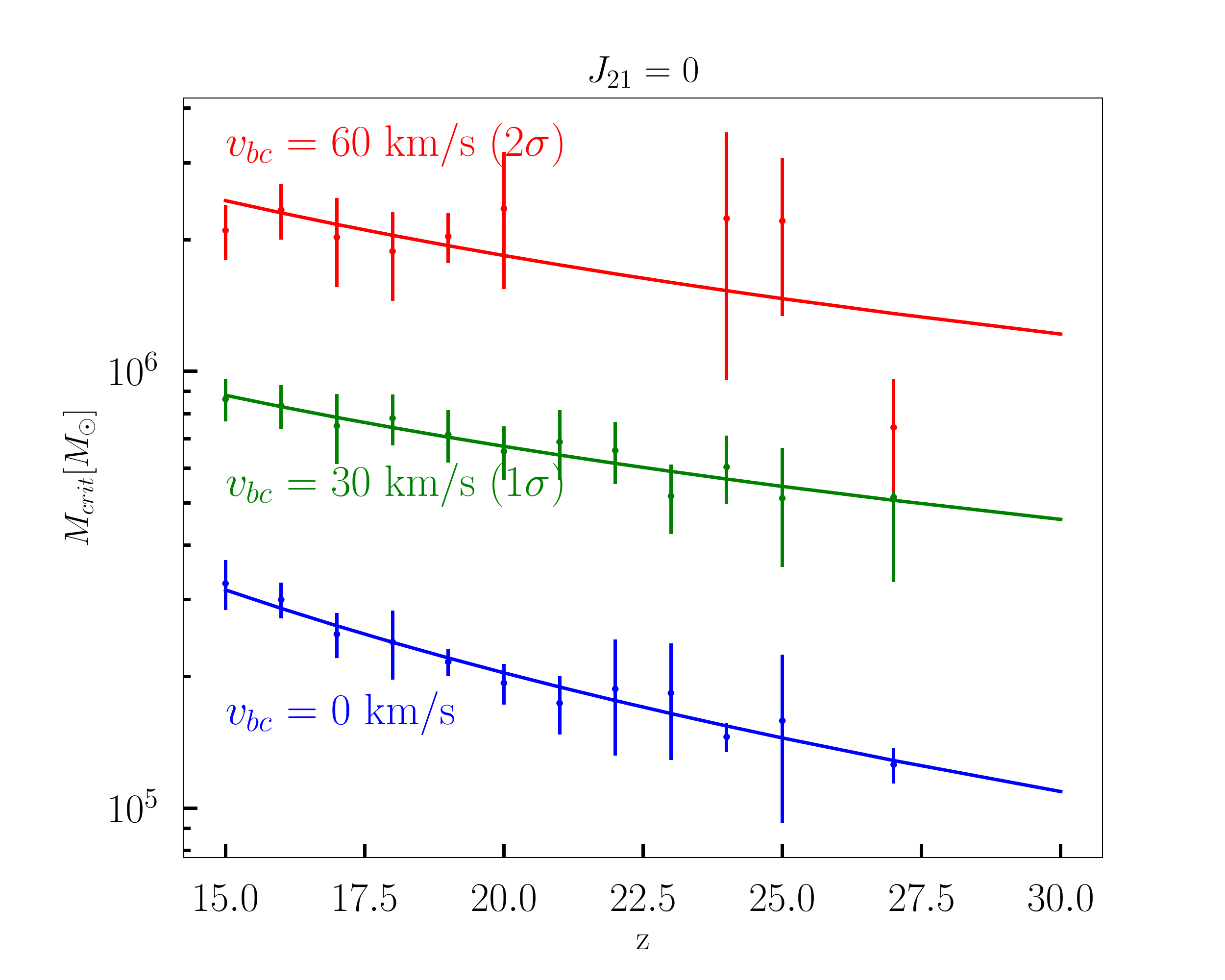}
    \caption{$M_{\rm crit}$ as a function of redshift without a LW background. The three lines show results from simulations with three different dark matter-baryon streaming velocities at recombination: 0, 30 km/s (1$\sigma$) and 60 km/s (2$\sigma$). Halos in regions with high streaming velocity need to be more massive in order to host cold dense gas and hence have higher $M_{\rm crit}$. The $z$-dependence becomes somewhat less steep with increasing streaming velocity, which can be explained by the fact that streaming velocity decreases with time as (1+$z$).}
    \label{fig:J0_vbcs}
\end{figure}

Next, we look at the effect of baryon-dark matter streaming on \mcrit\ in the absence of a LW flux.
Figure~\ref{fig:J0_vbcs} shows $M_{\rm crit}$ as a function of redshift without any LW background present. The three lines show $M_{\rm crit}$ for 3 different dark matter-baryon streaming values at recombination, corresponding to no streaming, 30 km/s ($1\sigma$) and 60 km/s (2$\sigma$).
In the regions with high streaming velocity, dark matter halos need to be more massive, with deeper potential wells, to have sufficient dense gas at their center, resulting in increased \mcrit.
From Table~\ref{tab:fits}, we can see that \mcrit\ increases by nearly a factor of 3 for the case with $1\sigma$ streaming and by nearly a factor of 10 for the case with $2\sigma$ at $z = 20$.

Figure~\ref{fig:J0_vbcs} shows that the $z$-dependence of \mcrit\ does not change as prominently as for the case with a LW background. From Table~\ref{tab:fits}, the exponent $\alpha$ characterizing the $z$-dependence decreases from 1.61 for no streaming to 0.99 for streaming of 30 km/s at recombination ($1\sigma$) and to 1.06 for streaming of 60 km/s at recombination ($2\sigma$). 
As can be seen from Figure~\ref{fig:J0_vbcs}, for $2\sigma$ streaming, \mcrit\ becomes nearly $z$-independent if we exclude the point for $z = 27$. This is not surprising given the fact that the streaming velocity decreases as $\propto (1+z)$ and so is more effective at suppressing the build up of baryons at high redshifts. Hence, $M_{\rm crit}$ increases more at high redshifts as compared to low redshifts, resulting in a shallower slope.

\subsection{Combined LW Flux and Streaming Velocities}

\begin{figure}
    \centering
    \includegraphics[width=0.5\textwidth]{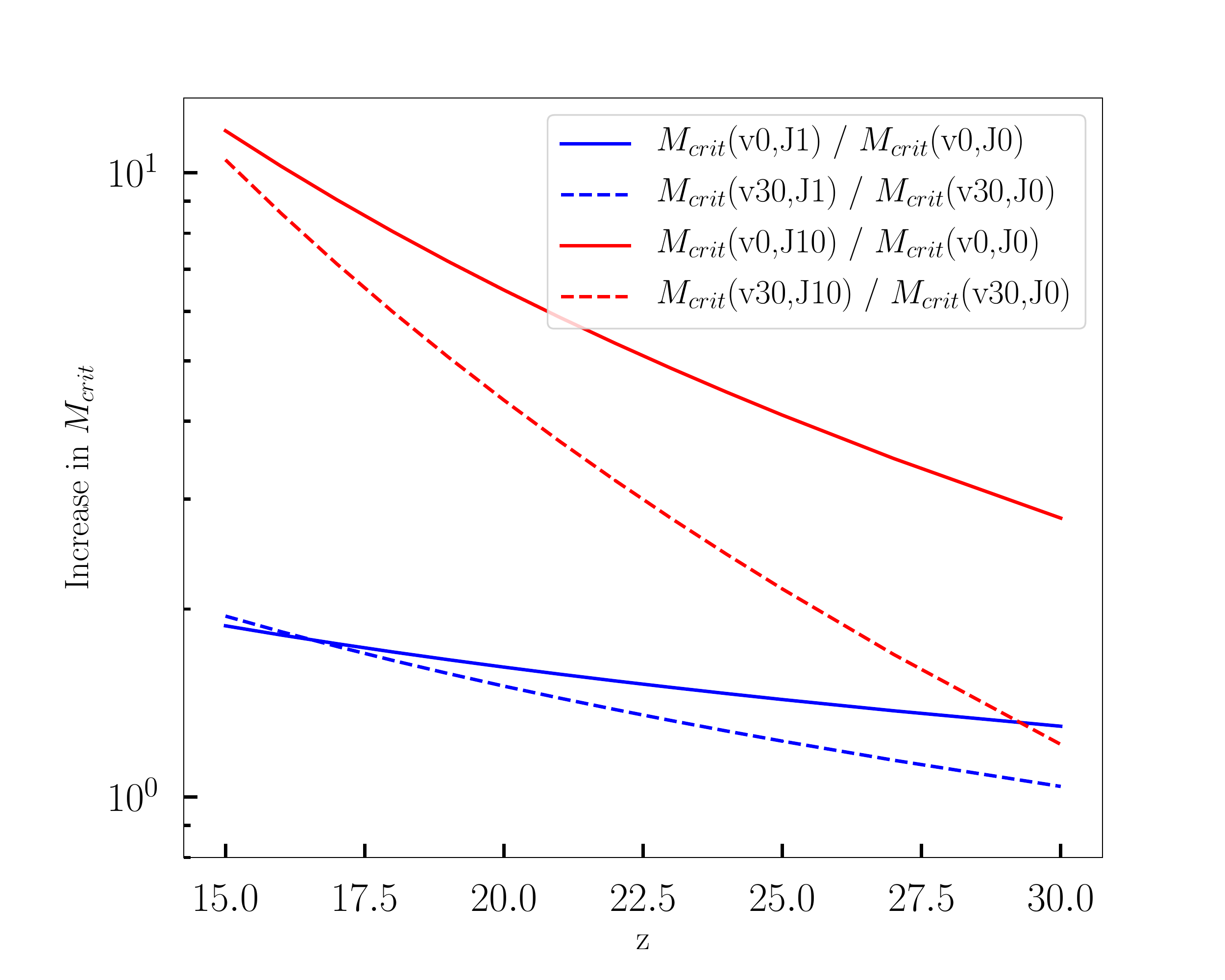}
    \caption{The increase in $M_{\rm crit}$ when a LW background is added (as a ratio), either in the presence or absence of dark matter-baryon streaming. The blue and red lines denote ratios of $M_{\rm crit}$ when $J_{\rm LW}$ is increased from $J_{\rm LW}=0$ to 1 and from 0 to 10 $J_{21}$, respectively. The solid lines show the respective ratios when there is no streaming present, whereas dashed lines correspond to the case where the streaming velocity is 30 km/s ($1\sigma$) at recombination. If the effects of LW background and streaming on $M_{\rm crit}$ were independent and multiplicative, then the solid and dashed lines would overlap each other in both cases. Instead we see that the increase in $M_{\rm crit}$ because of a LW background radiation is less prominent if you happen to be in a region with high streaming velocity. }
    \label{fig:increase_in_mcrit}
\end{figure}

We now look at the most general case, when both a LW background and dark matter-baryon streaming are present. 
\cite{Fialkov13} assumed that the effects of these processes are independent of each other and that the increase in $M_{\rm crit}$ from both processes would be multiplicative. As we have run simulations with multiple values of streaming and LW background present simultaneously, we can test this underlying assumption.

Figure~\ref{fig:increase_in_mcrit} shows the increase in $M_{\rm crit}$ as a ratio when the LW background is increased from $J_{\rm LW} = 0$ to $J_{21}$ (blue) and from 0 to $10 J_{21}$ (red), keeping the streaming the same. The solid lines show this increase for a region with no baryon-dark matter streaming, whereas dashed lines show the increase for a region corresponding to a streaming velocity of 30 km/s ($1\sigma$) at recombination. If the effects due to streaming and LW background were completely independent of each other and were multiplicative in nature, we would expect the solid and dashed lines to overlap. The fact that the dashed line is lower than the solid line for most of the redshift range suggests that the increase in $M_{\rm crit}$ because of LW flux is lower if it happens to be in a region with high streaming. In other words, the two effects are not entirely independent and the combined impact is not fully multiplicative -- instead, the combination of both tends to be slightly less effective than if they were each operating independently.

As noted above, the $z$-dependence of $M_{\rm crit}$ becomes steeper with increasing LW background, as shown by the $z$-dependence of the ratios of $M_{\rm crit}$ in Figure~\ref{fig:increase_in_mcrit}. This phenomenon is stronger in the region with baryon-dark matter streaming, which can be seen from the fact that the $z$-dependence of the $M_{\rm crit}$ ratios is steeper for the region with streaming present (dashed line) than without (solid line).

\begin{figure*}
    \centering
    \includegraphics[width=\textwidth]{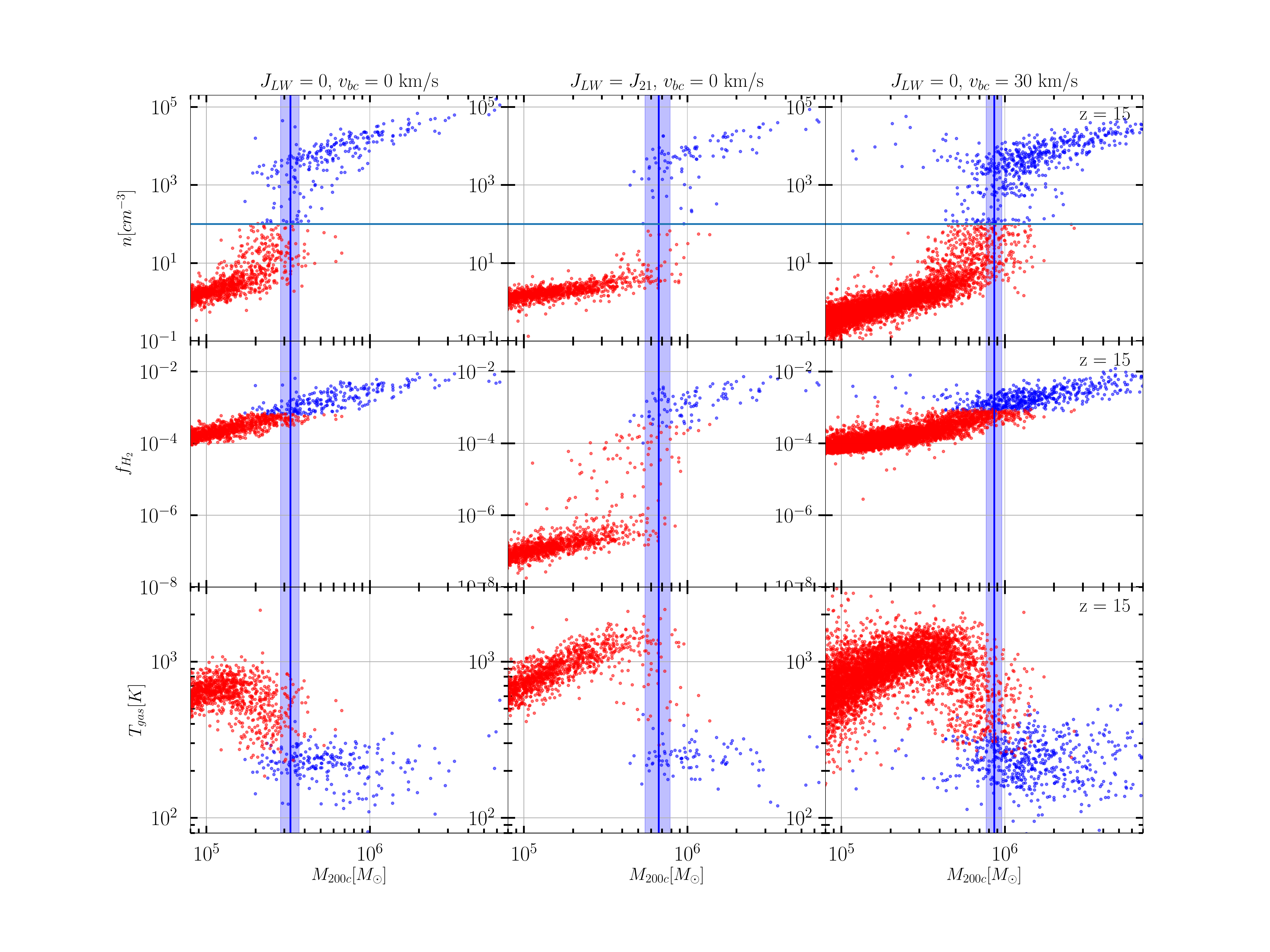}
    \caption{The gas number density (top panels), molecular hydrogen fraction (middle panels) and temperature (bottom panels) for the densest cell in each halo at $z = 15$. The three columns represent cases with no LW flux and no streaming (left), LW flux but no streaming (center) and streaming but no LW flux (right). The blue dots represent halos that have cold dense gas and red dots represent halos without cold dense gas. The vertical blue line and the shaded region around it denotes \mcrit\ for each of the three runs, and our estimate of the uncertainty on it.}
    \label{fig:central}
\end{figure*}

\subsection{Gas properties of the central regions of halos}
\label{subsec:gas_properties}
Finally, we present the gas properties of the central regions of the halos. Figure~\ref{fig:central} shows the number density, molecular hydrogen fraction, and gas temperature for the densest cell in each dark matter halo as a function of virial mass for three simulations at $z = 15$. The left column shows a simulation with neither LW flux nor baryon-dark matter streaming; the middle column represents a run with $J_{\rm LW} = J_{21}$ but no streaming, while the right column presents a case with a streaming velocity ($v_{\rm bc}$) of 30 km/s ($1\sigma$) (at recombination) in the absence of any LW flux. The blue dots represent halos that have at least one cell with cold dense gas and red dots represent halos without cold dense gas. The vertical blue lines and the shaded region around them indicate the critical mass for each run (and the uncertainty of that measurement, as described above). We can see that all of these properties show trends with the virial mass, but with a significant scatter.

The top row of Figure~\ref{fig:central} shows that for warm halos (those without cold-dense gas, shown as red points), the gas density increases steadily with the virial mass (or temperature) of the halo. This is in contradiction with the common assumption \cite[e.g.,][]{Trenti09} that the gas density in a halo is assumed to be only a function of redshift (this assumption is based on the idea that the central density is a fixed multiple of the mean density at that redshift). Instead, we find that the density increases nearly linearly with halo mass. Part of this increase is due to the entropy of the gas that arises due to heating at high redshift from the CMB background -- this entropy ($K \propto T/n^{2/3}$) can be higher than the entropy due to shock heating from gas falling into the dark matter halo, and therefore results in an enhanced pressure which resists compression. If (as we see below) $T \sim T_{\rm vir} \sim M_{\rm vir}^{2/3}$, then, for fixed $K$, $n \propto M_{\rm vir}$, as observed in \cite{Visbal14_nogo}.

This approximately linear scaling for the warm halos persists for different mass ranges in the three columns, depending on the cooling properties of the gas. In each case, near \mcrit, the cooling time of the central gas becomes shorter than the Hubble time and the gas starts to cool and increase in density up to nearly $\SI{1e4}{cm^{-3}} (n_{crit})$ and then increases slowly with virial mass for more massive halos (although the central density for halos with cold-dense gas may depend on the artificial pressure support described above). 

The gas temperature also shows a transition near \mcrit\ in the bottom row of Figure~\ref{fig:central}. For warm halos, $T_{gas} \sim T_{\rm vir}$, as expected for gas which is in virial equilibrium; however as cooling becomes efficient, the gas temperature drops significantly and rapidly, from $T_{\rm vir}$ down to about 200 K for halos with cold-dense gas (below this temperature, the $H_2$ cooling becomes increasing inefficient).

The middle row of Figure~\ref{fig:central} shows the molecular hydrogen fraction ($f_{H_2}$) as a function of virial mass. For the cases without LW flux (left and right columns), $f_{H_2}$ show a very clear monotonic trend with virial mass -- in particular, despite the rapid change in density and temperature, there is no break near \mcrit. The middle panel, with a LW background of $J_{21}$, has a lower molecular hydrogen fraction for warm halos, as expected. This gas is in photo-dissociative equilibrium, and the slow increase in $f_{H_2}$ arises from the temperature dependence of the formation rate. Self-shielding of gas from the LW background starts to become important near \mcrit\ resulting in a rapid increase of $f_{H_2}$, essentially up to a fraction consistent with the other simulations (that have no LW background). The effect of self-shielding is discussed further in Section~\ref{sec:discussion}.

\subsection{Scatter on \mcrit}
\label{subsec:scatter}

\begin{figure}
    \centering
    \includegraphics[width=0.45\textwidth]{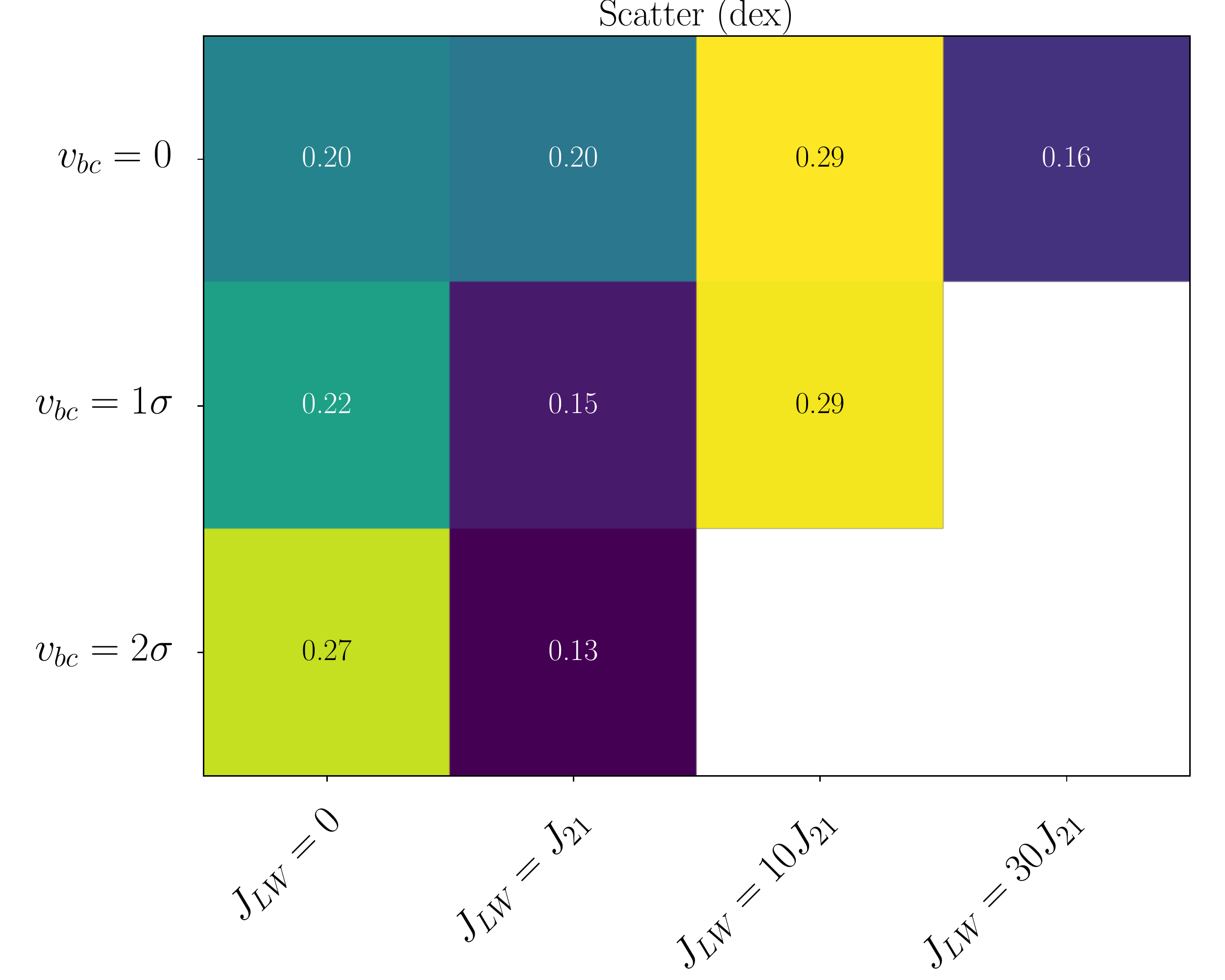}
    \caption{Relative scatter in dex (difference between $Q_3$ and $Q_1$) on \mcrit. We expect it to be converged for the cases with a large number of halos with cold dense gas (e.g., for $v_{\rm bc} = 0$, $J_{\rm LW} = 0, 1$; $v_{\rm bc} = 30$ km/s, $J_{\rm LW} = 0$ cases) and to be a lower limit of the underlying scatter for other cases. }
    \label{fig:scatter}
\end{figure}

The transition from warm halos to halos with cold dense gas does not happen abruptly at \mcrit\ and there is a significant scatter around it. We try to quantify this scatter by finding the masses corresponding to bins where 25\% and 75\% of the halos have cold dense gas in Table~\ref{tab:fits} (denoted $Q_1$ and $Q_3$, respectively). Figure~\ref{fig:scatter} shows the difference between $Q_3$ and $Q_1$ in log space. From this figure, we see the scatter is in the 0.2-0.3 dex range, with no clear trend with baryon-dark matter streaming or radiation strength.

The scatter is accurately measured when there is a large number of halos with cold dense gas. With fewer halos, the transition between halos with warm and cold dense gas is more likely to be abrupt, resulting in a smaller scatter, as is the case for e.g. $v_{\rm bc} = 0$; $J_{\rm LW} = 30 J_{21}$ or $v_{\rm bc} = 2\sigma$; $J_{\rm LW} = J_{21}$. To verify this claim, we split the halos for the $v_{\rm bc} = 0$; $J_{\rm LW} = 0$ case (which has many halos) into 6 samples of equal size and calculate the scatter on each of them. The values of the scatter on the split samples are $\sim 0.1$, which is smaller than the scatter of 0.2 when including all halos. For $v_{\rm bc} = 0$, the scatter increases from $J_{\rm LW} = J_{21}$ to $10 J_{21}$, as the number of cold dense halos increase as we shift from a box size of 0.5 h$^{-1}$ Mpc to 1 h$^{-1}$ Mpc. We conclude that the underlying scatter is equal to or larger than the scatter reported here. We believe that the estimated scatter is converged to the underlying scatter when the uncertainty on \mcrit\ (i.e. the bin sizes used) is much smaller than the value of \mcrit\ estimated (e.g., for $v_{\rm bc} = 0$, $J_{\rm LW} = 0, 1$; $v_{\rm bc} = 30$ km/s, $J_{\rm LW} = 0$ cases).

\section{Discussion}
\label{sec:discussion}

\subsection{An Emperical Fit for $M_{\rm crit}(J_{\rm LW}, v_{\rm bc}, z)$}
\label{subsec:fit}

One of the key aims of this work is to give a simple fit for $M_{\rm crit}(J_{\rm LW}, v_{\rm bc}, z)$ based on our simulation suite, which can then be used in various analytic models. In this section, we provide fits at a few different levels of accuracy and leave it to the reader to decide if they want to use the simple fit we provide or if they prefer to use a better fitting function based on the individual $M_{\rm crit}$ values we report.

Using the method described in Section~\ref{sec:method}, we calculate the critical mass for a given LW background, streaming velocity, and for a specific redshift. We fit for the redshift evolution with the following simple form:
\begin{equation}
    M_{\rm crit}(z) = M_{z=20} \cdot \left(\frac{1+z}{21}\right)^{-\alpha}.
\end{equation}
All of the fits are done using the Scipy function `curve\_fit'.
This provides us with two fit parameters: $M_{\rm crit}$ for redshift 20 $M_{z=20}$ and the redshift exponent $\alpha$, along with their uncertainties. 
`curve\_fit' uses $\chi^2$ minimization for fitting and calculating the uncertainty using our measured values of \mcrit.
We adopt $z=20$ as our pivot point because it is the center of our range.

As mentioned in Section~\ref{sec:results}, when we vary the LW background and baryon-dark matter streaming, we find that this changes the normalization ($M_{z=20}$) as well as the slope ($\alpha$). Hence we need to provide both of these parameters as a function of LW background and streaming.

\begin{equation}
    M_{\rm crit}(J_{\rm LW}, v_{\rm bc}, z) = M_{z=20}(J_{\rm LW}, v_{\rm bc}) \cdot \left(\frac{1+z}{21} \right)^{-\alpha(J_{\rm LW}, v_{\rm bc})}.
\end{equation}

We assume a simple functional form for $M_{z=20}$ and $\alpha$ as follows:

\begin{equation}
\begin{aligned}
    M_{z=20}(J_{\rm LW}, v_{\rm bc}) = (M_{z=20})_0\cdot \left(1+J_{\rm LW}/J_0 \right)^{\beta_1}\\
    \cdot \left(1+v_{\rm bc}/v_0 \right)^{\beta_2} \cdot
    \left(1+J_{\rm LW} v_{\rm bc}/Jv_0 \right)^{\beta_3}
\end{aligned}
\end{equation}
and 
\begin{equation}
\begin{aligned}
    \alpha(J_{\rm LW}, v_{\rm bc})) = \alpha_0\cdot \left(1+J_{\rm LW}/J_0 \right)^{\gamma_1} \cdot \left(1+v_{\rm bc}/v_0 \right)^{\gamma_2} \\
    \cdot \left(1+J_{\rm LW} v_{\rm bc}/Jv_0 \right)^{\gamma_3}.
\end{aligned}
\end{equation}
We have assumed one term each for the LW background and streaming dependence and one term for the cross-dependence.
However, if we simply fit these expressions with the constraints, we find there are too many parameters with too few data points, resulting in degeneracies between $J_0$ and $\beta_1$, $v_0$ and $\beta_2$. Therefore, we fix the pivot points $J_0$, $v_0$ and $Jv_0$ to be 1, 30 and 3 respectively and fit for the slopes $\beta$'s. $M_{z=20}$ was fit in the log space, whereas $\alpha$ was fit in the linear space. The fitted $\beta$'s and fit values for the overall amplitude, $M_{z=20}$ are provided here:
\begin{align}
    (M_{z=20})_0 &= \num{1.96e5} \pm \num{1.33e4} M_{\odot}, \\
    \beta_1 &= 0.80 \pm 0.06, \\
    \beta_2 &= 1.83 \pm 0.14, \\
    \beta_3 &= -0.06 \pm 0.04.
\end{align}

The fitted parameters for $\alpha$ are as follows:
\begin{align}
    \alpha_0 &= 1.64 \pm 0.11,\\
    \gamma_1 &= 0.36 \pm 0.03,\\
    \gamma_2 &= -0.62 \pm 0.15,\\
    \gamma_3 &= 0.13 \pm 0.03.
\end{align}

In addition to this global fit, we also provide in Table~\ref{tab:fits} fits for the individual simulations (i.e.~$M_{z=20}$ and $\alpha$ for all the combinations of LW background radiation and streaming velocities with the appropriate errors). We also provide \mcrit\ with uncertainties for all redshifts for all cases in an accompanying file. Users can fit it with a different fitting function of their choice if they prefer.
 
The uncertainty on \mcrit\ we provide depends on the number of halos with cold dense gas in the simulation. Therefore we have a smaller uncertainty on \mcrit\ for lower LW flux and streaming. Apart from providing \mcrit\ corresponding to a mass bin with half of the halos with cold dense gas, we also provide an estimate on the scatter on it. Columns 5 and 6 of Table~\ref{tab:fits} provide the halo masses corresponding to bins that have 25\% and 75\% of the halos with cold dense gas, respectively, for $z = 20$. The scatter estimate is useful for semi-analytic models that populate dark matter halos with first stars.

\subsection{A simple model for explaining the $z$-dependence}
\label{subsec:model}
A simple analytical model such as those described in \cite{Machacek01} or \cite{Trenti09} can be used to explain some of the redshift trends we see in terms of the quantities at the central regions of the halos. 

The condition for collapse can be defined as the cooling time being shorter than the Hubble time at that redshift. The cooling time is given as
\begin{equation}
    t_{\rm cool} = \frac{1.5 n k_B T_{\rm vir}}{\Lambda(T_{\rm vir}) n_H n_{H_2}},
    \label{eq:cooling_time}
\end{equation}
where we have assumed the gas temperature to be equal to the virial temperature of the halo. The cooling function behaves as $\Lambda (T) \propto T^{3.4}$ for the temperature between \SI{120}{K} and \SI{6400}{K}. Hence the cooling time of the gas varies as $t_{\rm cool} \propto T^{-2.4} n_{H_2}^{-1}$. 

For the case with no LW flux and no streaming velocity, the central molecular hydrogen density can be fit as a power-law function of $T_{\rm vir}$ and $z$ as
\begin{equation}
    n_{H_2} \propto (1+z)^{1.62} T_{\rm vir}^{2.0}. 
\end{equation}
This is an approximate scaling relation from our simulations for halos without cold dense gas (e.g. red points in Figure~\ref{fig:central}). To get this relation, we first fit $n_{H_2}$ as a function of $T_{\rm vir}$ at a given redshift and then fit for a $z$-dependence of $n_{H_2}$ at $T_{\rm vir} = \SI{1000}{K}$.
Using this relation and the evolution of Hubble time as $t_H \propto (1+z)^{-1.5}$, we can find a redshift dependence of the critical virial temperature. As $M_{\rm vir} \propto T_{\rm vir}^{3/2} (1+z)^{-3/2}$, we get $M_{crit} \propto (1+z)^{-1.54}$ which is very close to our measured $z$-dependence of $(1+z)^{-1.58}$.

In the presence of LW flux, the molecular hydrogen density increases rapidly with increasing redshift. For $J_{\rm LW} = 10 J_{21}$, 
\begin{equation}
    n_{H_2} \propto (1+z)^{5.58} T_{\rm vir}^{2.2}.
\end{equation}
Using this relation in equation~\ref{eq:cooling_time} and equating it to the Hubble time gives a $z$-dependence of $M_{\rm crit} \propto (1+z)^{-2.83}$. This does not match exactly with the observed $z$-dependence of $\propto (1+z)^{-3.74}$, although it follows the qualitative trend of steeper $z$-dependence than the case without LW flux. Our simple analytic model does not precisely explain the steepening of the $z$-dependence with increasing LW flux. One possible reason for this discrepancy could be that the simple power-law fitting formula we used for $n_{H_2}$ does not capture its dependence on $T_{\rm vir}$ and $z$ accurately. The exponent for the relation between $n_{H_2} - T_{\rm vir}$ changes as a function of redshift as well, but we have not included it so as to keep the fitting formula relatively simple.

The steeping of the $z$- dependence of \mcrit\ in the presence of LW flux can be primarily attributed to a steeper $z$-dependence  of the central molecular hydrogen density. At high redshifts, gas densities are higher which results in more effective self-shielding. This explains why the molecular hydrogen density increases rapidly with redshift in the presence of LW flux.
We find that the molecular hydrogen density is in equilibrium in presence of LW flux. We conclude this by comparing the self-shielding factor ($f_{sh}$) calculated by assuming an equilibrium for molecular hydrogen density and the self-shielding factor ($f_{sh}$) used in the simulation (see also the discussion in Section~\ref{subsec:self-shielding}) and finding them nearly equal. This justifies the steeper slope of the molecular hydrogen with redshift in presence of LW flux.

\subsection{Comparison with previous works}

In this subsection, we compare our results with previous works. 
\cite{Tegmark97} estimated a minimum mass for forming first stars using a simple analytic model. They compared the cooling time of the gas in a halo as a function of gas density, temperature and molecular hydrogen fraction, with the Hubble time in order to estimate its fate. Their estimated minimum mass depends more strongly on redshift than we find in our simulations and increases with decreasing redshift (see Figure 6 in \cite{Tegmark97}).
\cite{Tegmark97} find an increase in \mcrit\ from \SI{1e5}{M_{\odot}} to \SI{1e7}{M_{\odot}} approximately from redshift of 80 to 15 in the absence of LW radiation. We, on the other hand, find \mcrit\ to be below \SI{3e5}{M_{\odot}} until redshift $z=15$.

\cite{Haiman2000} estimated the critical flux ($J_{\rm crit}$) above which the star formation in a halo is prevented as a function of $T_{\rm vir}$ and $z$ (Figure 6).
They report that halos cannot cool for a virial temperature below \SI{e2.4}{K}. This corresponds to \mcrit\ changing with redshift as $\propto (1+z)^{-1.5}$ which matches nicely with the $z$-dependence we see for the case without any LW flux ($\propto (1+z)^{-1.58}$), although \mcrit\ corresponding to a virial temperature of \SI{e2.4}{K} at $z = 20$ is \SI{9.7e5}{M_{\odot}}, which is significantly higher compared to our estimate of \SI{2.05e5}{M_{\odot}}. 

To estimate the redshift evolution of \mcrit\ in the presence of LW flux in \cite{Haiman2000}, we compare the virial temperatures corresponding to redshift 10 and 20 for $J_{\rm LW} = 10^{-2} J_{21}$. From $z = 10$ to $z = 20$, $T_{\rm vir,crit}$ changes from \SI{1e3.6}{K} to \SI{1e2.8}{K}. This corresponds to a $z$-dependence of the critical virial temperature as $\propto (1+z)^{-2.8}$. Because $M_{\rm vir} \propto T_{\rm vir}^{3/2} (1+z)^{-3/2}$, this corresponds to $M_{\rm crit} \propto (1+z)^{-5.7}$ which is somewhat steeper than our results, although it does match our finding that the LW flux tends to steepen the dependence of \mcrit\ on redshift. We note that their \mcrit\ for $J_{\rm LW} = 0.1 J_{21}$ at $z = 20$ corresponds to \SI{6.1e6}{M_{\odot}}, which is again higher than the value we find. 

\cite{Machacek01} used a statistical sample of halos to find a minimum mass for halos to form cold-dens gas. They divided halos into two redshift bins with $z > 24$ and $z < 24$ and found no evidence of epoch dependence in them. We find a strong redshift dependence in our results; however it is not clear if there is a significant difference due to the smaller number of halos they were able to analyze. Otherwise, their no LW background results agree well with what we find here; however, their LW dependence is much stronger than found in this paper, which we ascribe to their neglect of H$_2$ self-shielding.

\cite{Trenti09} used a similar analytic model to estimate the critical halo mass to form Pop III stars. They assume $T_{\rm gas} = T_{\rm vir}$, $n \propto (1+z)^3$ and a power-law relation between $f_{H_2}$ and $T_{\rm vir}$. In the absence of LW radiation, a minimum mass can be derived by equating the cooling time of the gas to the Hubble time. This minimum mass is given by $M_{t_H-cool} \approx \SI{1.54e5}{M_{\odot}} (\frac{1+z}{31})^{-2.074}$. In the presence of a LW background, a similar minimum mass can be found by assuming an equilibrium for the formation and dissociation of molecular hydrogen to estimate the molecular hydrogen fraction. This minimum mass is given by $M_{H_2-cool} \approx \SI{6.44e6}{M_{\odot}} J_{21}^{0.457}(\frac{1+z}{31})^{-3.557}$. \cite{Trenti09} argue that for a dark matter halo to be able to cool via H$_2$, its mass must be above both of these limits. 

The analytic model in \cite{Trenti09} predicts a few properties of $M_{\rm crit}$ that we see in our simulations. It predicts a redshift evolution of $M_{\rm crit}$ in a qualitatively similar way for the case without LW radiation. It also predicts a steeper $z$-dependence when LW radiation is present; however it does not predict a $z$-dependence that changes with the background LW radiation as we see in our simulations. For the case without LW flux, \cite{Trenti09} estimate \mcrit\ of $\sim \SI{3.3e5}{M_{\odot}}$ at $z = 20$, which is close to our estimated value of $\SI{2.05e5}{M_{\odot}}$. For the case with $J_{\rm LW} = J_{21}$, \cite{Trenti09} estimate \mcrit\ to be $\sim \SI{2.57e7}{M_{\odot}}$ at $z = 20$, which is much higher than our estimated value of $\SI{3.31e5}{M_{\odot}}$.
We speculate that this is in part due to their assumption of a central density that depends only on redshift and not halo mass, which differs significantly from what we saw in Section~\ref{subsec:gas_properties}.

\cite{Schauer19} studied the effect of the streaming velocity on $M_{\rm crit}$ of a large statistical sample using the moving mesh code AREPO (but without any LW background). 
We compare our results for \mcrit\ with $M_{halo, 50\%}$ quoted in \cite{Schauer19} which is defined as the average halo mass above which 50\% of the halos contain cold gas. This is similar in spirit to the criterion assumed in this work, but quantitatively distinct.
Our results are in broad agreement, with streaming leading to a factor $\sim 3$ ($\sim 10$) increase when a velocity of 1$\sigma$ (2$\sigma)$ was adopted. In detail, there are a few important differences.
First, \cite{Schauer19} found a constant $M_{halo, 50\%}$ as a function of redshift (with or without baryon-dark matter streaming), whereas we find that \mcrit\ tends to increase with decreasing redshift, consistent with a fixed virial temperature for the case without baryon-dark matter streaming. 
Second, we note that their $M_{halo, 50\%}$ corresponding to the case without LW background or streaming is about $\SI{1.6e6}{M_{\odot}}$, significantly larger than our $M_{\rm crit}$, which we find to be close to $\SI{2e5}{M_{\odot}}$. They also report the minimum halo mass to contain cold-dense gas, and this is lower, but still higher than our \mcrit\ value, indicating that the differences cannot be solely due to different definitions. More work is required to understand these differences.

During the final stages of the completion of this paper, a related study investigating the effect of LW flux and baryon-dark matter streaming velocity was released \citep{Schauer20}, building on the results just discussed. Here we briefly compare our results. Again, the broad qualitative picture is in agreement, with a LW background leading to an increase in the minimum mass for cold-dense gas formation, but some important differences in the details.
\cite{Schauer20} probe a somewhat different parameter space than us, in particular lower LW backgrounds (0.01 $J_{21}$ and 0.1 $J_{21}$), which makes a direct comparison difficult. If the fit from \cite{Schauer20} for the $J_{\rm LW}$ dependence on \mcrit\ is extrapolated to $J_{\rm LW} = J_{21}$, that corresponds to an increase in \mcrit\ by nearly a factor of 10 when $J_{\rm LW}$ changes from 0 to $J_{21}$, whereas we find an increase by just a factor of two.

\cite{Skinner20} also model Pop III star forming in low mass halos including self-shielding. Although, it is difficult to do a one-to-one comparison with their results, \cite{Skinner20} find Pop III stars forming in halos of mass \SI{3e5}{M_{\odot}} (and larger) at $z = 20$ for a LW background of slightly less than $J_{21}$ which is broadly consistent with our results. 

\subsection{Resolution tests}

We have performed a number of tests to check for the convergence of our simulations, both with respect to the dark matter particle mass (which is set by the initial grid resolution in \textsc{enzo}) as well as the number of AMR levels which determine the minimum size of the baryon cells at high densities. 

\begin{figure}
    \centering
    \includegraphics[width=0.5\textwidth]{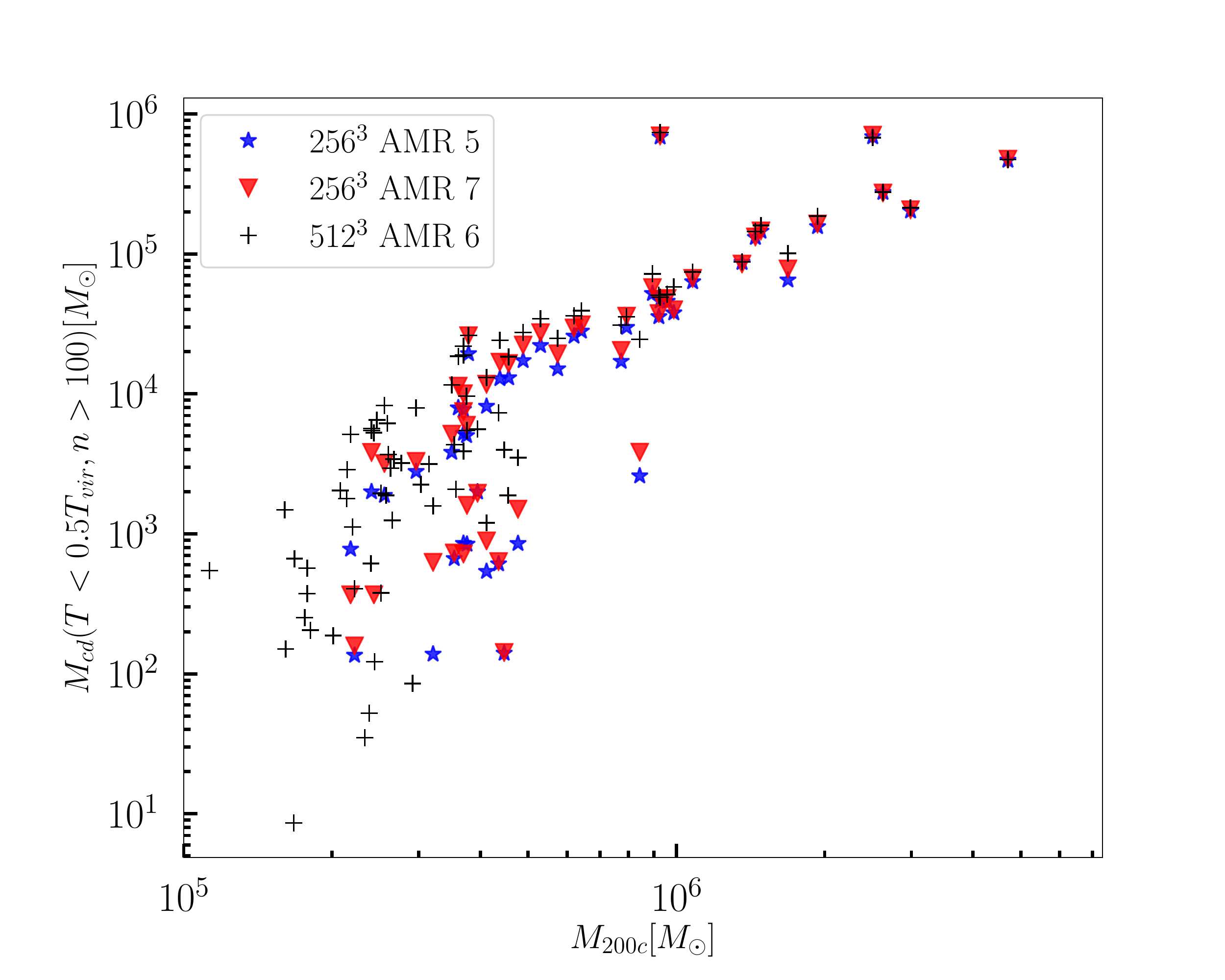}
    \caption{The effect of resolution on the total mass of cold dense gas in three simulations with the same initial conditions but different resolutions. We plot $M_{200c}$ of each halo identified on the $x$-axis and the mass of the cold dense gas ($T < 0.5 T_{\rm vir}$, $n > \SI{100}{cm^{-3}}$) on $y$-axis for $z = 22$ for a box of size 0.5 h$^{-1}$ Mpc. The blue stars and red triangles correspond to a $256^3$ grid with two different AMR levels (see text for additional details). The black plus symbols correspond to $512^3$ grid with the same cell resolution as the run with red triangles.}
    \label{fig:resolution}
\end{figure}

Figure~\ref{fig:resolution} shows the total mass of the cold dense gas  ($T < 0.5 T_{\rm vir}$, $n > \SI{100}{cm^{-3}}$) as a function of halo mass at $z = 22$ for a 0.5 h$^{-1}$ Mpc box. The symbols denote three different resolutions. The black plus symbols represent a simulation with a $512^3$ grid and 6 AMR levels, which corresponds to the dark matter particle mass of $\sim 100 M_{\odot}$ and smallest cell size of $\sim \SI{21.8}{cpc}$. The red triangles correspond to a $256^3$ grid with a dark matter particle mass of $\sim 800 M_{\odot}$ and the same cell size. Finally, the blue stars correspond to a $256^3$ grid with only 5 AMR levels which corresponds to a minimum cell size of $\sim \SI{85}{cpc}$.

We can see from Figure~\ref{fig:resolution} that most of the blue stars and red triangles have overlap; in other words, they have similar amounts of cold dense gas, particularly for the more massive halos. In most of our simulations we use the higher spatial resolution of the two runs analyzed here. When we compare red triangles and black plus symbols, which have the same spatial resolution but different dark matter particle masses, we see more black plus symbols near the low mass end representing more halos with cold dense gas. This is not surprising as a $\SI{e5}{M_{\odot}}$ halo would be resolved by just over 100 dark matter particles for the lower resolution, whereas it needs to be resolved by at least 500 particles to have an accurate estimate of the gas content \citep{Naoz09}. Hence, for the runs with low LW background and streaming, we use a resolution corresponding to a dark matter particle mass of $\SI{100}{M_{\odot}}$ such that even a halo of mass $\SI{e5}{M_{\odot}}$ is resolved by 1000 particles. However for halos of mass $\SI{e6}{M_{\odot}}$ or more, we can see that black plus symbols and red triangles overlap, as they are well resolved by the low resolution simulation as well. Therefore, in order to have more large halos for runs with high LW background and streaming where \mcrit\ would be higher, we use a larger 1 h$^{-1}$ Mpc box with a $512^3$ grid, corresponding to a dark matter particle mass of 800 $M_{\odot}$.

\subsection{Self-shielding}
\label{subsec:self-shielding}

Many previous works investigating the effect of LW radiation on \mcrit\ have ignored the self-shielding of gas from LW radiation \citep{Machacek01,Wise-Abel07,Oshea07}. \cite{Wolcott11, Wolcott19} have emphasized the importance of self-shielding and provided a prescription for including it in simulations using a multiplicative factor $f_{sh}$ which depends mainly on the column density of molecular hydrogen and gas temperature.
Because of self-shielding, the effective LW radiation penetrating the central dense region is lowered and the central region can have higher molecular hydrogen densities. In the presence of self-shielding, we expect \mcrit\ to decrease for a fixed LW background, as lower mass halos would be able to have sufficient molecular hydrogen to cool.

\begin{figure}
    \centering
    \includegraphics[width=0.5\textwidth]{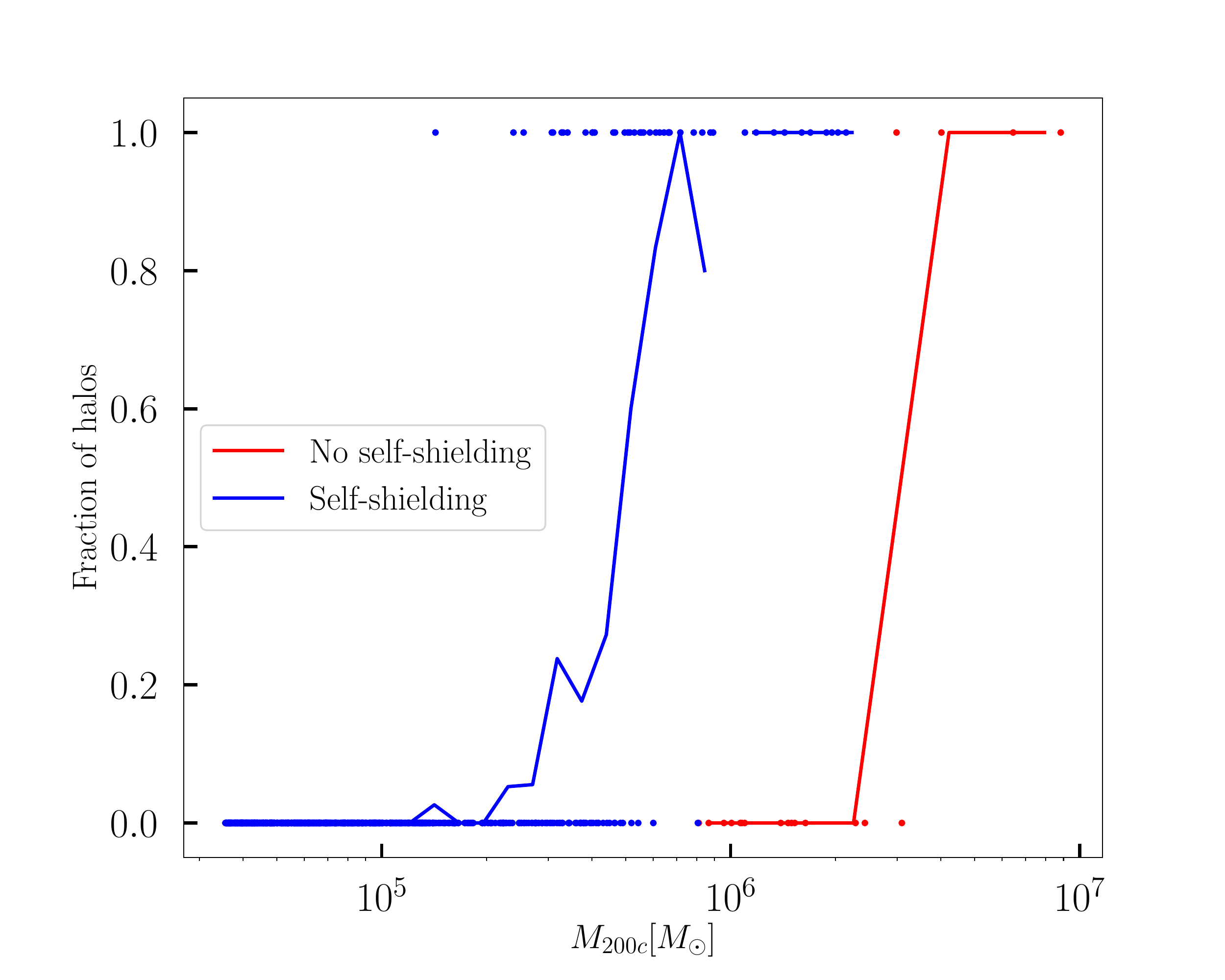}
    \caption{Comparison of two simulations, both with $J_{\rm LW} = J_{21}$ at $z = 20$, but one (red dots) without self-shielding from LW radiation, and the other (blue dots) with a self-shielding prescription based on \cite{Wolcott19}. Each dot indicates a halo, with the x-axis indicating the halo mass and the y-axis showing the presence ($y=1)$ or absence ($y=0$) of cold-dense gas. The solid lines show the fraction of halos that have cold dense gas at a given mass bin using the method described in Section~\ref{sec:method}. Inclusion of self-shielding decreases \mcrit\ significantly. }
    \label{fig:selfshielding}
\end{figure}

Figure~\ref{fig:selfshielding} compares two simulations with (blue) and without (red) self-shielding for $J_{\rm LW} = J_{21}$ at $z = 20$. The red and blue dots represent individual halos and have a $y$-value of 1 if they have cold dense gas and 0 if they do not. The solid lines denote the fraction of halos that have cold dense gas in each mass bin as explained in Section~\ref{sec:results}. The simulation with self-shielding has significantly smaller \mcrit\ as can be seen by comparing the blue and red solid lines in this figure.

\begin{figure}
    \centering
    \includegraphics[width=0.45\textwidth]{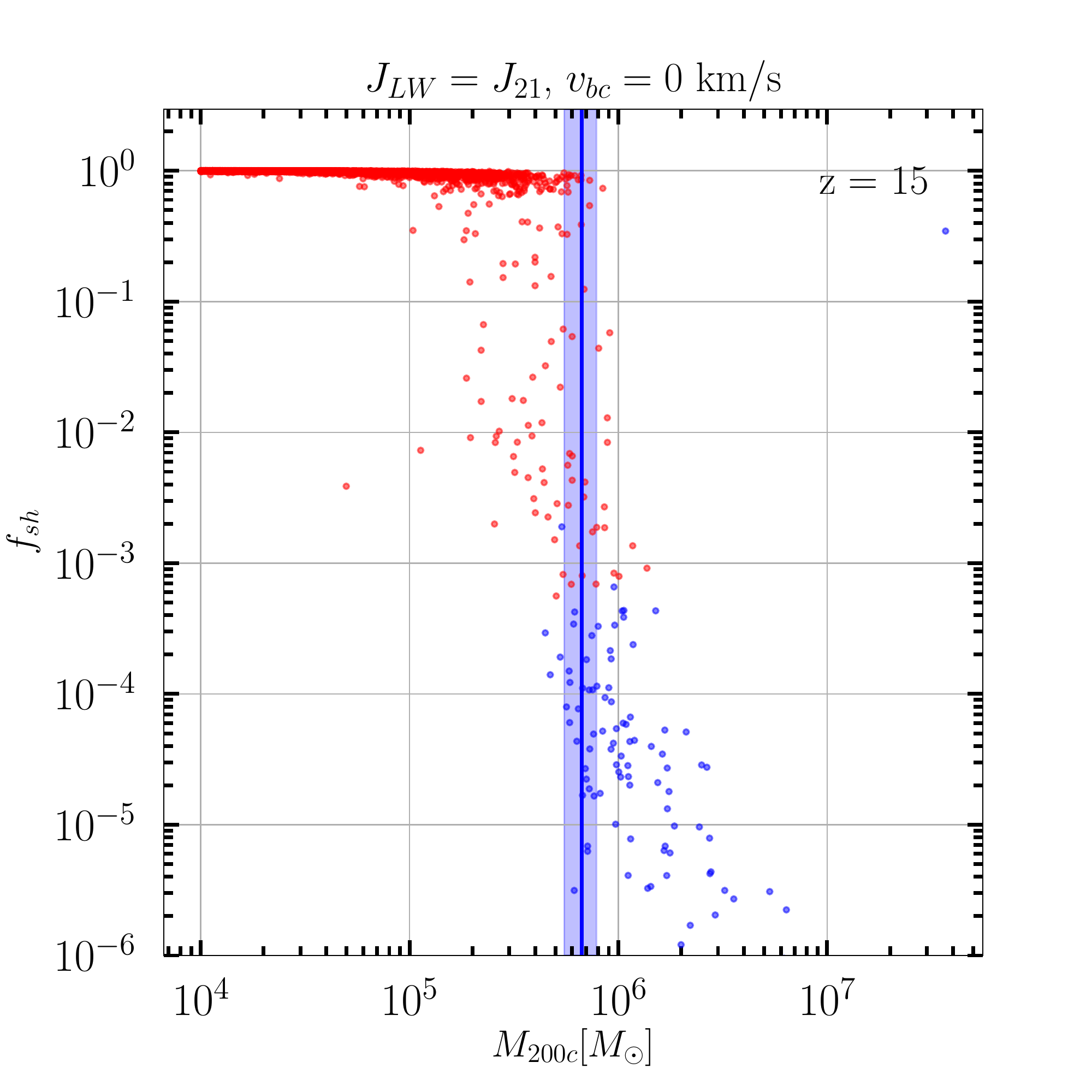}
    \caption{The self-shielding factor $f_{sh}$ as defined in \cite{Wolcott19} for the densest cell of each halo for the simulation with no streaming and a LW background of $J_{21}$. The effective LW radiation is equal to the background LW flux multiplied by $f_{sh}$. The blue (red) dots represent halos that have at least one cell with (without) cold dense gas. The vertical blue line and the shaded region around it shows \mcrit\ and its uncertainty. Halos with cold dense gas have effectively self-shielded themselves from the background LW radiation.}
    \label{fig:fsh}
\end{figure}

To explicitly demonstrate the importance of self-shielding, in Figure~\ref{fig:fsh} we show the self-shielding parameter $f_{sh}$ as a function of halo mass for a simulation with $J_{\rm LW} = J_{21}$ and no streaming at $z = 15$.  Low mass halos have no self-shielding and have $f_{sh} = 1$. Halos with mass closer to \mcrit\ have higher gas density, resulting in a reduced $f_{sh}$ which further increases the molecular hydrogen density and enhances cooling. As shown in Figure~\ref{fig:fsh}, warm halos with masses slightly smaller than \mcrit\ also have self-shielding factors which are significantly lower than 1. In the absence of self-shielding, molecular hydrogen densities would be sufficiently high only in more massive halos, resulting in an increased \mcrit.

\section{Summary and Conclusion}
\label{sec:summary}
We performed a simulation suite of minihalos (halos with virial temperatures below the atomic cooling limit) using the cosmological hydrodynamics code \textsc{enzo} in order to constrain the dependence of the critical dark matter halo mass for Pop III star formation on LW radiation, baryon-dark matter streaming, and redshift. We performed a set of simulations varying the LW background and the streaming velocity over the expected range for each candidate, using simulation volumes large enough to create a large sample of such halos with a resolution sufficient to accurately identify when cooling produces cool, dense gas. We then analyzed the simulations to determine \mcrit, the dark matter halo mass for which 50\% of halos hosted cold-dense (potentially star-forming) gas. We also report the uncertainty on \mcrit\ and the scatter (the mass range over which 25\%-75\% of the halos have cold-dense gas).
Our conclusions can be summarized as follows.

\begin{enumerate}
    \item We identify a clear redshift dependence of the critical mass, finding that \mcrit\ increases with decreasing redshift as $\propto (1+z)^{-1.58}$ for the case with no LW flux and no streaming, consistent with a fixed virial temperature, a relation which has been predicted in analytic models but not previously seen in numerical works with a statistically significant sample of halos.
    
    \item We find a LW background increases \mcrit\ and that the redshift dependence of \mcrit\ changes from a slope of $-1.58$ to $-5.70$ as the LW flux increases from 0 to $30 J_{21}$ and becomes shallower with increasing dark matter-baryon streaming velocity. 
    \item We find that self-shielding of the gas from LW radiation is important and decreases \mcrit\ for a given LW flux. \mcrit\ increases by a factor of 2 when going from $J_{\rm LW}$ of 0 to $J_{21}$ -- nearly an order of magnitude smaller increase than previous results that did not incorporate self-shielding.
    
    \item We performed simulations in which both LW flux and streaming are present in order to critically examine the idea that their impact on \mcrit\ are independent of each other and act in a multiplicative way. We conclude that the two effects are not entirely independent and the combined impact is not fully multiplicative -- instead, their impact in combination tends to be somewhat less effective than if they were each operating independently. The increase in \mcrit\ can be smaller by nearly a factor of 2-3 at high redshifts when compared to an estimate based on the assumption of independence.
    
    \item We provide a fit for \mcrit\ as a function of LW flux, baryon-dark matter streaming and redshift which can be used by semi-analytic models to make predictions for Pop III stars and their observable signatures.
\end{enumerate}

\acknowledgments{
We thank Zoltan Haiman for useful discussions. The computations in this paper were carried out on the Rusty supercomputer of the Flatiron Institute. EV acknowledges support from NSF grant AST-2009309. GLB acknowledges support from NSF grants AST-1615955 and OAC-1835509 and NASA grant NNX15AB20G.  We gratefully recognize computational resources provided by NSF XSEDE through grant number TGMCA99S024, the NASA High-End Computing Program through the NASA Advanced Supercomputing Division at Ames Research Center, Columbia University, and the Flatiron Institute. This work made significant use of many open source software packages. These are products of collaborative effort by many independent developers from numerous institutions around the world. Their commitment to open science has helped make this work possible.}

%



\software{\textsc{astropy} \citep{2013A&A...558A..33A},  
          \textsc{yt} \citep{yt},
          \textsc{enzo} \citep{Enzo, Enzo_joss}
          }




\bibliography{mihir_pop3}{}
\bibliographystyle{aasjournal}



\end{document}